# Economy-based Content Replication for Peering Content Delivery Networks

Al-Mukaddim Khan Pathan, Rajkumar Buyya, James Broberg* and Kris Bubendorfer**

| | | |
|---|---|---|
| **Gri**d Computing and **D**istributed **S**ystems (GRIDS) Laboratory<br>Department of Computer Science and Software Engineering<br>The University of Melbourne, Australia<br>*{apathan,raj}@csse.unimelb.edu.au* | *Distributed Systems and Networking (DSN) Laboratory<br>School of Computer Science and Information Technology<br>RMIT University, Australia<br>*jbroberg@cs.rmit.edu.au* | **School of Mathematics, Statistics and Computer Science<br>Victoria University of Wellington<br>PO Box 600, Wellington 6001, New Zealand<br>*kris@mcs.vuw.ac.nz* |

**Abstract:** Existing Content Delivery Networks (CDNs) exhibit the nature of closed delivery networks which do not cooperate with other CDNs and in practice, *islands* of CDNs are formed. The logical separation between contents and services in this context results in two content networking domains. In addition to that, meeting the Quality of Service requirements of users according to negotiated Service Level Agreement is crucial for a CDN. Present trends in content networks and content networking capabilities give rise to the interest in interconnecting content networks. Hence, in this paper, we present an open, scalable, and Service-Oriented Architecture (SOA)-based system that assist the creation of open Content and Service Delivery Networks (CSDNs), which scale and supports sharing of resources through peering with other CSDNs. To encourage resource sharing and peering arrangements between different CDN providers at global level, we propose using market-based models by introducing an economy-based strategy for content replication.

## 1. Introduction

Content Delivery Networks (CDNs) [1], which evolved first in 1998, replicate content over several mirrored Web servers, strategically placed at various locations around the globe to deal with *flash crowds* and to enhance response time. A CDN has some combination of a content-delivery infrastructure, a request-routing infrastructure, a distribution infrastructure and an accounting infrastructure. CDNs improve network performance by maximizing bandwidth, improving accessibility and maintaining correctness through content replication. In a typical content delivery environment Web server clusters are located at the edge of the network to which the end-users are connected. A content provider can sign up with a CDN provider for service and have its content placed on the content servers. The content is replicated either on-demand when users request for it, or it can be replicated beforehand, by pushing the content to the surrogate servers. A user is served with the content from the nearby replicated Web server. Thus the users end up communicating with a replicated CDN server close to them and retrieves files from that server.

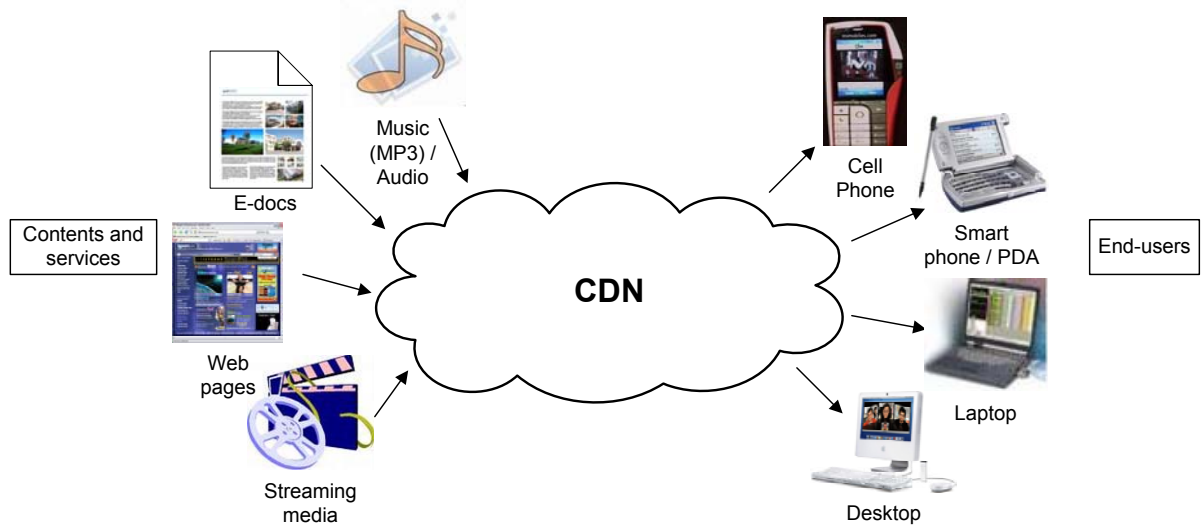

Figure 1: Content/services provided by a CDN

CDN providers ensure the fast delivery of any digital content. They host third-party content including static content (e.g. Static HTML pages, images, documents, software patches etc), streaming media (e.g. audio, real time video etc) and varying content services (e.g. directory service, e-commerce service, file transfer service etc.). The sources of content are large enterprises, Web service providers, media companies, news broadcasters etc. The end-users interact with the CDN specifying the content/service request through cell phone, smart phone/PDA, laptop, desktop etc. Figure 1 depicts the different content/services served by the CDN to end-users.

## 1.1. Motivation and scope

Existing Content Delivery Networks (CDN) are proprietary in nature – individual companies own and operate them – they comprise their own closed delivery networks, which are expensive to set up and maintain. They also have limited scalability. Running a global CDN is even more costly, requiring an enormous amount of capital and labor. In addition, content providers typically subscribe to one CDN provider and thus can not use the resources of multiple CDNs at the same time. Such closed and non-cooperative model results in "islands" of CDNs. Moreover, the logical separation between contents and services under the "content delivery and distribution" and "content services" domains is undesirable considering the ongoing trend in content networking. A unified network that supports coordinated composition and delivery of content and services would be much better.

Furthermore, commercial CDNs make specific commitments to their consumers by signing a Service Level Agreement (SLA). It describes provider's commitment and specifies penalties if those commitments are not met. So, if a particular CDN provider is unable to provide quality service to the end-user requests, it may result in SLA violation and end up costing the provider. To cut expenses and to avoid adverse business impact, CDN providers should partner together.

In this paper, we present a model for an open, scalable and Service-Oriented Architecture (SOA)-based system. This system helps to create open Content and Service Delivery Networks (CSDNs) [3] that scale well and share resources with other CSDNs through cooperation and coordination, thus overcoming the island CDN problem. To encourage resource sharing and peering arrangements between different CDN providers at global level, we propose using market-based models in resource allocation and management inspired from their successful utilization in the management of autonomous resources, especially in global grids [4]. Hence, we introduce an economy-based content replication strategy that involves on-demand placement of outsourced content in the surrogates of peering CDNs. The use of economic mechanisms in this context has following benefits:

- We propose a Virtual Organization model for peering the CDNs. A VO consists of real CDN providers, which are self-interested and autonomous stake-holders. Hence, an economic model is suitable to represent this scenario and to regulate interactions among the participants. The problem can be manageable in this way through analyzing emergent marketplace behavior.
- A peering environment of content networks is highly dynamic in nature, where the availability of resources changes over time. In this environment, an economic model is appropriate to exploit the dynamism of the market to make more informed decisions on the fly.
- An economic model may be the basis of a self-regulating replication strategy that dynamically adapts to the changes in the end-user request patterns.

The rest of the paper is organized as follows: Section 2 establishes the significance and relevance of our proposal; Section 3 addresses the shortcomings of related work; Section 4 presents the proposed model for peering CDNs; Section 5 enlightens the economy-based model for content replication; Section 6 formulates the system model for auction; and Section 7 provides a summary with expected contributions and future directions.

## 2. VO-based peering of CDNs

## 2.1. VO concepts

Virtual organizations [17] are composed of a number of semi-independent autonomous entities (including different individuals, departments, and organizations) that come together to share resources and to collaborate on shared goal(s). These entities co-exist, cooperate and sometimes compete with one another in a virtual marketplace. One or more entities in such virtual marketplace may sometimes realize the potential benefit of collaborating with other entities by selling and/or exchanging resources. When such potential is recognized, relevant entities go though a process of forming a new VO to exploit it. Each involving entity attempts to attract the attention of potential customers and interacts with them by advertising the cost and quality of its services, with a goal of selling them in such a way that maximizes individual gain. The participants of a VO must cooperate and coordinate their activities in order to effectively manage the VO and to

Page 2 of 13

achieve the common goal(s). In dynamic environment, a VO may no longer be available due to the change in context at any time. It will then need to either disband or re-arrange itself into a new organization through re-negotiation that better fits the enduring circumstances.

When a VO is formed, a number of issues are taken into account, including:

- An entity responsible for forming a VO should be able to recognize circumstances in which it should initiate VO formation.
- A participating entity while joining in a VO should be able to determine the conditions under which it is profitable for it to join.
- Given a number of offers, the entity that initiates VO formation should be able to determine the best offer(s) in terms of economic benefit.
- In support to these issues, full-bodied description of quality of service should be present within the VO in order to determine the extent to which services meet customers' requirements.

## 2.2. VO formation scenarios

In our approach, a Virtual Organization (VO) is formed through coordination of Web servers of different CDNs who have come together to share resources and collaborate on common goal(s). A VO of CDN providers is interoperable in the sense that peering relationship within a VO can be initiated among arbitrary parties, accommodating new participants dynamically with changes in context. A VO in the peering CDN environment may vary in terms of purpose, scope, size, and duration. VOs in such environment are of two types: *short-term on-demand VOs* and *long-term VOs* with established SLAs.

A short-term VO is formed for short time duration, based on current user request pattern to prevent the generation of *hotspots*. Consider the following scenario as motivation for short-term VO formation. Suppose that the content of www.cnn.com is hosted by the CDN provider Akamai [5][6]. The popularity of this Web site is high as it provides exclusive coverage of the latest news around the world. Akamai's Web servers receive significant user requests to serve the latest content on behalf of www.cnn.com. A sudden news outburst (e.g. 9/11 incident in USA [7], death of 'Crocodile Hunter' in Australia [8]), demanding to the end-users of a part of the world may cause heavy workload on Akamai's Point-of-Presences (POPs) in that particular region. As a result a *hotspot* can be generated. It will cause Akamai's POPs in that region to be unable to cope with the strain. Eventually the Web servers will be totally overwhelmed with the sudden increase in traffic, and CNN's Web site will be temporarily unavailable. Such sudden spike in Web content requests is termed as *flash crowd* [2] or *SlashDot effect* [9]. Web servers of a particular CDN provider experience such sudden outburst of load due to the content providers with extensive Web presence.

In the peering CDN environment, generation of hotspot due to flash crowd can be resolved through the formation of short-term VOs. A short-term VO intervenes with sudden spike in requests for particular Web content(s), which results in heavy workload on certain Web server(s) of a particular CDN. Hence, surrogates of peering CDNs form a goal-oriented constellation of distributed semi-autonomous entities and excess load is distributed to the less loaded Web servers of other CDNs. Such peering arrangement should be automated within computationally reasonable time frame to exploit the evolved niche. A short-term VO is formed on-demand and the policy for such VO formation is established responsively to handle the evolved situation. It is phased out when the workload returns to normal.

On the other hand, a long-term VO is formed for events which may be known in advance. In long-term VOs, CDN providers collaborate for longer period of time. Formation of long-term VO exemplifies the existence of established policies and negotiated SLAs among the participating entities. To better understand the formation of long-term VO, consider the following scenario. Suppose that the ICC Cricket World Cup 2007 is being held in the Caribbean, and www.cricinfo.com is supposed to provide live media coverage. As a content provider, www.cricinfo.com has an exclusive SLA with the CDN provider, Akamai [5][6]. However, Akamai doesn't have a point of presence (POP) in Trinidad and Tobago (a Caribbean island), where most of the cricket matches will be held. As being the host of most of the cricket matches, people of this particular part of Caribbean are expected to have enormous interest in the live coverage provided by www.cricinfo.com. Since Akamai is expected to be aware of such event well in advance, its management can take necessary initiatives to deal with the evolving situation. In order to provide better service to the users, Akamai management might decide to place its surrogates in Trinidad and Tobago, or they might use their other distant edge servers. Firstly, placing new surrogates just for one particular event would be costly and might not be useful after the event. On the other hand, Akamai risks its reputation if it can't provide quality service according to client requests, which could violate the SLA and still cause profit reduction. Hence, the solution for Akamai could involve partnering with other CDN provider(s) to form a VO in order to deliver the service that it could not provide otherwise. A Long-term VO remains for the duration of the event. Automation for long-term VO formation is not a must since such situation is known before-hand.

Thus, by collaborating with other CDN providers though the formation of VO, content networks can better satisfy their customers and meet their QoS requirements.

## 2.3. Research issues

In this section, we present the unique issues that are to be addressed for peering CDNs.

*Load Distribution*: The crucial question to be addressed for distributing loads among peering CDNs is:
- How to ensure reduced bandwidth consumption (less traffic that needs to go over the network) and reduced server load (fewer requests for a server to handle)?

*Coordination of CDNs*: Related to this issue, the following question is to be addressed:
- What kind of coordination mechanisms need to be in place those ensure effectiveness, and allow scalability and growth of cooperative CDNs?

*Service and policy management*: In this field, the following questions need to be addressed:
- How to make value-added services an infrastructure service that is accessible to the customers?
- What types of Service Level Agreements (SLA) are to be negotiated among CDN participants? What policies can be generated to enforce SLA negotiation? How these policies can be managed?

*Pricing of contents and services*: The following question is to be addressed in this context:
- How do CDN providers achieve maximum profit in a competitive environment, yet maintaining the equilibrium of supply and demand?

## 3. Related work

Peering of Content Delivery Networks is gaining popularity among researchers of the scientific community. Several projects/works are being conducted for finding ways to peer the CDNs for better overall performance. In this section, we outline the efforts for internetworking of Content Delivery Networks:

The internet draft by IETF proposes a Content Distribution Internetworking (CDI) Model [10], which allows the CDNs to have a means of affiliating their delivery and distribution infrastructure with other CDNs who have content to distribute. According to the CDI model, each content network treats neighboring content networks as *black boxes*, which uses commonly defined protocol for content internetworking, while internally uses its proprietary protocol. The CDI Internet draft assume a federation of CDNs but it is not clear how this federation is built and by which relationships it is characterized. It also does not address the issue of policy management for enforcing SLAs among the internetworked CDNs.

An architecture for Content Distribution Internetworking (CDI) is presented in [11]. It discusses the design, implementation and evaluation of a protocols architecture that can effectively support the interoperation and cooperation of separately administered CDNs. A CDI allows every CDN in that CDI both to augment the number of potential content consumers and to get the content access faster. This work shows that P2P models are not suitable for CDI construction since it does not provide significant benefits. Hence, some semi-centralized approach based on a star topology is used where an authoritative CDN is responsible for a particular group of content requests and the request is forwarded by this CDN to other CDN which will serve the requests. Thus, performance data has to pass only one hop since all CDNs forward their performance data to the CDNs to which they are federated. The protocol for CDI is termed as RIEPS (Routing IEP for Star topology). The main drawback of this protocol is – being a point-to-point protocol, if one end-point is down the connection remains interrupted until that end-point is restored.

The Content Internetworking Router (CiRouter) [12], implemented by FastTide allows clients to choose from where it wants to retrieve content. When CiRouter receives a request for an HTML page, it delivers the performance measurement and selection code to the client originating the request. The client runs the code performing the response time measurements, analyzes output to choose a CDN and reports the result to the CiRouter. Hence, CiRouter modifies URLs of the embedded objects of the HTML page at runtime, and delivers the modified content to the client from the selected (by client) CDN. The disadvantage of this approach is – client has to be modified to run the performance code provided by the CiRouter.

CDN brokering [13] allows one CDN to intelligently redirect clients dynamically to other CDNs in that domain. It uses techniques offered by the DNS to redirect client requests to the best CDN. The client DNS request is forwarded to the brokering CDN server (BDS) that is authoritative for a particular domain. This DNS-based system is called as Intelligent Domain Name Sever (IDNS). IDNS responds to DNS requests intelligently based on a dynamic, load-sensitive configuration rather than using static information. The drawback is that, mechanism for IDNS is proprietary in nature and might not be suitable for a generic CDI architecture.

Content Network Advertisement Protocol (CNAP) [14] is an advertisement protocol which is designed to facilitate the interconnection of separately administered CDNs. It is intended to communicate information for the purpose of performing request routing decisions between interconnected CDNs. CNAP is not a routing protocol but can be used to exchange information that may be used for inter-CN request-routing decisions. The Internet draft illustrating the CNAP does not specify the topology of the overlay network that the protocol requires.

## 4. The model for peering CDNs

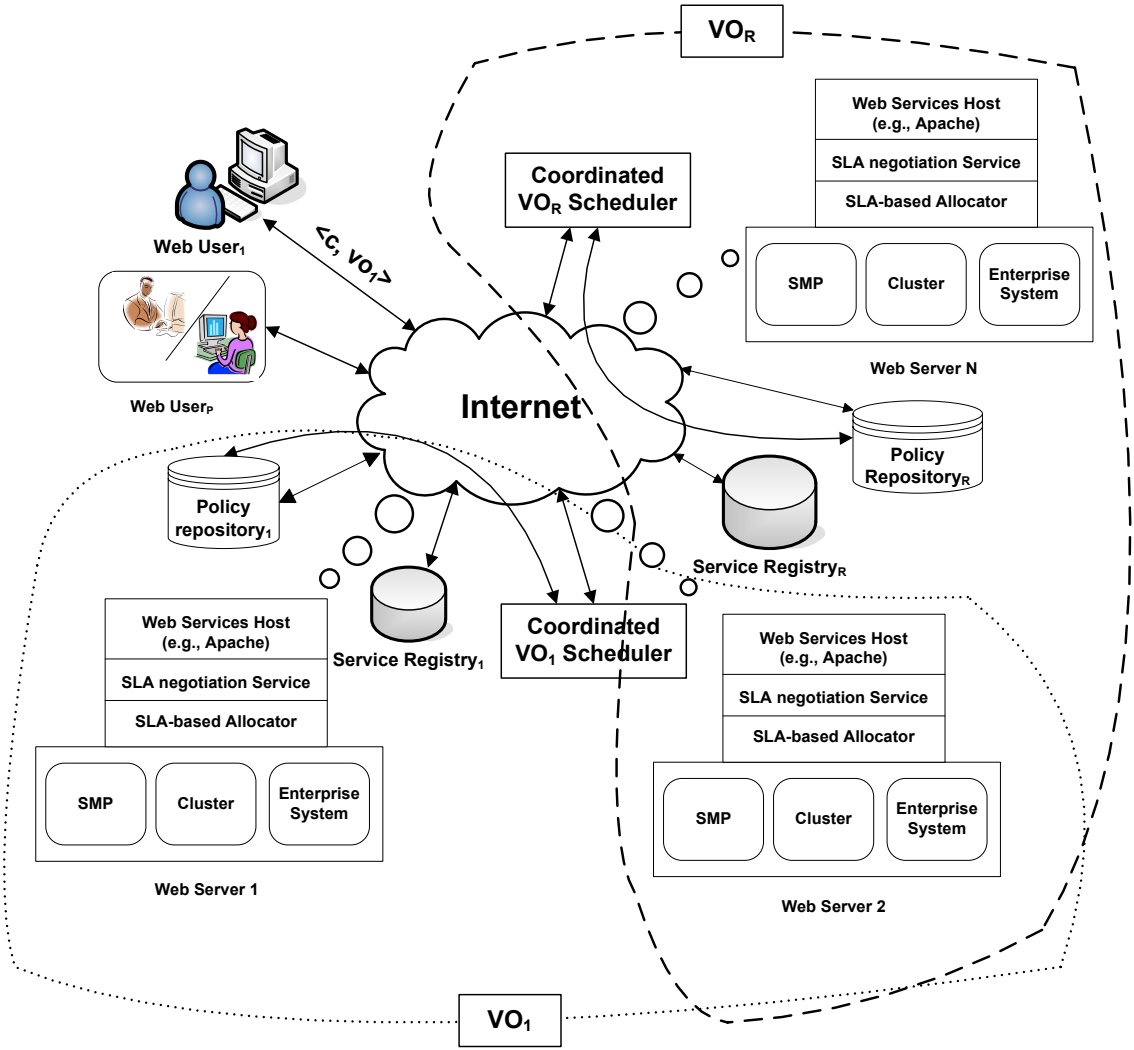

**Figure 2: Architecture of open, scalable, Service-Oriented Architecture (SOA)-based system to assist the creation of cooperative and coordinated CSDNs (Content and Service Delivery Networks)**

Our proposed system solves the *island* CDN problem, ensures the quality of services based on SLA negotiation, and solves the problem of logical separation in content networking domain. Architecture for such a system is shown in Figure 2. We propose a VO-based model for forming Content and Service Delivery Networks (CSDNs). In the proposed model, a VO is formed through coordination of Web servers of different CSDNs who have come together to share resources and collaborate on common goal(s). A VO of CSDN providers is interoperable in the sense that peering relationship within a VO can be initiated among arbitrary parties, accommodating new participants dynamically with changes in context. Web servers within each CSDN are capable of delivering services in order to meet the Quality of Service (QoS) requirements of the users. The VO-based model has mechanism to discover resources so that resource-sharing arrangements can be established and changed quickly. Hence, resources are discovered through the use of a service registry. The formation of VO may be stand alone or may be composed of a hierarchy of regional, national and international VOs. In this figure, each Web server represents a CSDN. A VO consists of CSDN Web servers, a coordinated VO scheduler, a service registry, and a policy repository. Web servers within each CSDN are capable of delivering services in order to meet the Quality of Service (QoS) requirements of the users. Users' interact transparently with the VO by requesting content from Web server(s) of a CSDN. A content request may initiate further VO activities (e.g. Inter-CSDN request routing, content replication and delivery in a peering arrangement), and thus the participating entities act as a single conceptual unit in the context of aimed services. Coordinated VO schedulers and service registries are autonomous components within the system which are dispersed at random, and they help in the formation of VOs of CSDNs. Policy repository is a VO-component owned by the CSDN which initiates a VO formation.

A brief description of the VO components is provided below:

**Web server –** The Web server is a CSDN's most important element. Servers are responsible for storing content and value-added services as infrastructure services and delivering them in a reliable and cooperative manner. Servers within each CSDN can deliver content and services to meet end-user QoS requirements. We can divide a server's structure into two layers: an overlay layer and the core. In the overlay layer, a server comprises a Web-service host (for example, Apache or Tomcat), an SLA-negotiation service module, and an SLA-based allocator. The negotiation module, with the help of a coordinated VO scheduler, is responsible for cooperating and coordinating with other servers (located in a local or global CSDN) through SLA-based negotiation. The SLA-based allocator delivers content and services based on the negotiated SLAs with other local or global CSDNs' servers. The Web server's core consists of high performance computing systems such as symmetric multiprocessors, cluster systems, or other enterprise systems (such as desktop grids). The server's underlying devices and tools must store the content and services and assist in responding quickly and reliably to client requests to meet the negotiated QoS requirements. For content and service location and routing, the Web servers' underlying technologies perform on-demand cooperative caching through coordination with other servers. Efficiently balancing the load across different Web servers is critical to produce the required QoS. Hence, servers are adopted with appropriate load- and resource-distribution strategies.

**Coordinated VO scheduler –** A coordinated VO scheduler is an autonomous entity that serves as a mediator for the peering CSDNs. In a peering CSDN environment, VO schedulers are dispersed at random. It is put in each VO and is responsible for ensuring collaboration and coordination with other CSDNs though policy exchange and scheduling of content and services. It also negotiates QoS parameters and resource allocation to maximize cooperative CSDNs' potential. When a VO is formed, the coordinated VO scheduler in it has the responsibility of ensuring VO activities in an effective manner. A key objective of the existence of VO scheduler within a VO is to ensure that the participating entities are able to adapt to changing circumstances (agility) and are able to achieve their objectives in a dynamic and uncertain environment (resilience). It also contributes in decision making for resource sharing, content replication, policy enforcement, and scheduling.

**Service registry –** A service registry enables CSDN providers to register and to publish their resources and service details. CSDN providers also flash out their resource requirements through service registry at times of need. For instance, in case of flash crowd, the convincible CSDN initiate VO formation by publishing its resource requirements through service registry. Thus, it takes the opportunity to seek resources from other peering CSDNs. An SLA-negotiator service and allocator module uses this service registry to discover potential CSDN providers to form VOs.

**Policy repository –** A policy repository stores the policies that the administrators generate. These policies are a set of rules to administer, manage, and control access to VO resources. They provide a way to consistently manage the components deploying complex technologies.

## 4.1. Policy management for enforcing SLAs

A policy in the context of peering CSDNs would be statements that are agreed upon by the participants within a VO. These statements define what type of contents and services can be moved out to a CSDN node, what resources can be shared between the VO participants, what measures are to be taken to ensure quality service based on negotiated SLA, and what type of programs/data must be executed at the origin servers. Within our proposed VO-model based CSDN architecture we apply the standard policy framework defined by the IETF/DMTF [15]. The basic policy framework is shown in Figure 3.

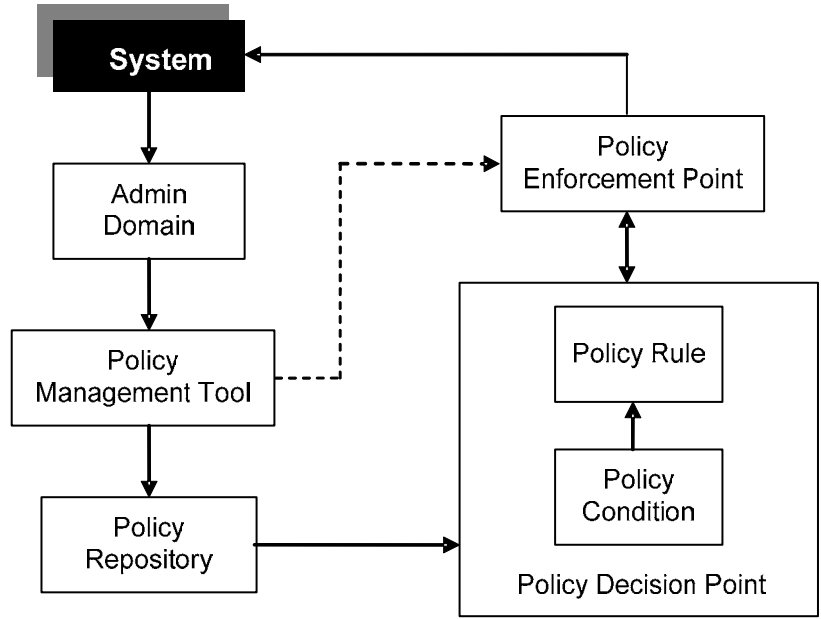

**Figure 3: Basic policy framework**

In the standard policy framework, the *admin domain* refers to an entity which administers, manage and control access resources within the *system* boundary. An administrator uses the *policy management tools* to define the policies to be enforced in the system. The *policy enforcement points (PEPs)* are logical entities within the system boundary, which are responsible for taking action to enforce the defined policies. The policies that the PEPs need to act on are stored in the policy repository. The results of actions performed by the PEPs have direct impact on the system itself. The *policy repository* stores polices generated by the administrators using the policy management tools. The *policy decision point* is responsible for retrieving policies from the policy repository, for interpreting them (based on *policy condition*), and for deciding on which set of policies are to be enforced (i.e. *policy rules*) by the PEPs.

The proposed model for CSDN in figure 2 can be mapped to the basic policy framework in Figure 3. The policy repository in figure 2 is responsible for storing policies generated by the policy management tool used by the VO administrator. The distribution network and the Web server components (i.e. SLA negotiation service, SLA based Allocator) are the instances of policy enforcement points (PEPs), which enforce the CSDN policies stored in the repository. Each VO schedulers are the instances of the policy decision points (PDPs), which determine the set of policies to be enforced at the time of coordination and cooperation among CSDNs. The policy management tool is administrator dependent and it is not shown in figure 2. A direct benefit of using such policy-based architecture is to reduce the operating cost of CSDNs and to meet end-user QoS requirements according to negotiated SLA.

## 5. The economic model

Given a set of properly placed surrogate servers in a VO-based peering CSDN infrastructure and a chosen content for delivery, a very important issue is the optimal placement of the outsourced content to the appropriate surrogate(s). Standardized economy concepts can be deployed for content replication within the structure of the VO model for CSDN. Here, we introduce a novel auction protocol for content replication in peering CSDNs environment. The mechanism underlying the auction protocol helps in deciding when and where to create and delete replicas. The aim is to minimize the network latency perceived by the user and to restrict the generation of Web 'hotspots' due to the excessive client requests for some content.The participation in a VO is due to profit motive. But such peering may result in free-riding where some CSDN providers use other's resources free of charge. Such free-riding can be avoided through using an auction model for replicating content among CSDN servers. The auction model should be able to provide incentives to all parties.

In our economy-based model, the goal of the auction protocol is to select the cheapest suitable Web server in order to replicate content there. Here we apply the buyer-driven auction mechanism, which is a type of Vickrey auction [16] . Vickrey auctions are second-price sealed-bid auctions with low messaging overhead, efficiency of allocations and lack of counter speculation. They involve a single negotiation round in which each bidder submits a bid to the auctioneer. Other bidders cannot see the bid. The bidding agent which makes the highest bid wins the auction but pays the price of the second-highest bid.

In our case, each CSDN provider is both a buyer and seller of its resources. A coordinated VO scheduler is the *auctioneer* in the peering CSDN environment that is responsible for holding auction within the VO. It starts an auction on behalf of a CSDN provider (i.e. buyer) for finding suitable surrogate server(s) in order to perform content replication. *Buyers* buy the storage space of Web server(s) of peering CSDNs in a particular region which incurs excessive client requests. *Sellers* are CSDN providers who sell the storage space of their Web server to the buyer. SLA-based allocator is the *bidding agent* that resides in each Web server. SLA negotiation module is the *communication mediator*, which is responsible for establishing and maintaining peering communications infrastructure. It also propagates auction messages between auctioneer and bidding agents.

The auctioneer starts an auction not for selling an item (in this case, storage space), but for buying it. Bidding agents bid with the price they are willing to sell the storage space of their Web servers. One bidder can not see the bid of other bidders. Auctioneer gathers bids from the bidding agents and selects the lowest bidding agent(s) as the winner and the winner is paid second-lowest bidding price by the auctioneer. In other words, our economic model uses a reverse Vickrey auction.

## 5.1. Formation of VO in the economic model

Having known the internals of the economic model, we now discuss the steps for formation of VO in our economic model:
1. A CSDN provider (buyer) realizes the need to replicate content to the surrogates of peering CSDNs. Buyer internally determines the maximum payable amount using the *Payoff Function* and announces its *Auction Policy*. Auctioneer starts auction on behalf of the buyer.
2. The bidding agent of seller (other CSDN providers) uses a *Bidding Function* to determine the bidding amount. A *Utility Function* is used to determine the potential economic benefit based on *expected revenue*.
3. Auctioneer collects bids from the bidding agents and selects the lowest bidding surrogate(s) as winner and winners are paid by the amount of second-lowest bid.
4. Hence, VO of buyer and seller CSDNs is formed and content is replicated to the winners' surrogates.
5. Re-negotiation through auction takes place when either of the following holds: (a) A seller varies its demand after winning; (b) Seller finds holding replicated content no longer economically beneficial for it; (c) A less demanding CSDN provider (except winner) comes up.

## 6. System model for auction

In our economic model, there is one buyer agent, one auctioneer who acts on behalf of the buyer and a fixed number of seller agents, who are the bidders or bidding agents. The auctioneer starts an auction process on behalf of a buyer agent that needs to buy storage space for replicating its content. Let us assume that $N$ denotes the set of CSDN providers where $i \in N$; $R$ denotes the set of client requests $r \in R$; $C$ denotes the set of all contents, $c \in C$. The $k$-th arriving request at time $t_k$ is $r_k$ – composed of $r_k^c$ and $r_k^l$, representing the content and the location components of the respective request. The storage requirement of content $c$ is defined as $S_c \geq 0$ (MB or GB). The network delay function $\sigma(r_k, i)$, specifies whether request $r_k$ is served by the provider $i$ within delay threshold, $D$; i.e., $\sigma(r_k, i) = 1$ if $\sigma(r_k, i) < D$; $\sigma(r_k, i) = 0$ otherwise. Each CSDN provider $i$'s incurred load at time $t_k$ is $L_i(t_k)$, which is the sum of all the loads imposed by each content request $r$, that is, $L_i(t_k) = \sum_k L_i^{r_k}(t_k)$. Now the client request cannot be handled by the CSDN provider $i$ either because of the network delay perceived by the user or because of provider's storage capacity constraints. It can be represented as:

$$\partial(L_i, t_k) = \left(\sum_k L_i^{r_k}(t_k)(1 - \sigma(r_k, i))\right) + \ell\left(\left(\sum_k L_i^{r_k}(t_k)\sigma(r_k, i)\right) - S_c\right)$$

where the first term is for user perceived network delay and the second term is for provider's storage capacity constraints. The constant $\ell$ is defined in such a way that the second term is cancelled out when it produces negative value. Hence,

$$\ell = \begin{cases} 1 & \textit{If } \left(\sum_k L_i^{r_k}(t_k)\sigma(r_k, i) - S_c\right) > 0 \\ 0 & \textit{Otherwise} \end{cases}$$

When $\partial(L_i, t_k) > 0$, provider (buyer) $i$ realizes monetary penalty or loss goodwill due to not satisfying customer QoS requirements. Hence, it decides to replicate content $c$ by announcing its requirements as *Auction Policy (AP)* in the service registry of peering CSDN environment. Hence, a coordinated VO scheduler (auctioneer) initiates auction on behalf of the buyer. The auction policy, $AP_c$, consists of:

*Service Requirements:*
- *Storage requirement:* The storage space required to replicate content $c$, defined as $S_c$ (MB or GB).
- *Upload rate:* Rate of transfer (kb/sec) to replicate content to the surrogate, denoted by $U_c$.
- *Download rate:* Rate of transfer (kb/sec) to serve user request from the replicated content, denoted by $D_c$.
- *Preferences:* The buyer's biasness for surrogate server(s) in a potential *hotspot* region.

*Duration:*
- *Time frame:* Time to hold the replica, denoted by $T$.

The bidding agent of seller (other CSDN providers) uses a *Bidding Function* to determine the bidding amount. A *Utility Function* is used to determine the potential economic benefit based on *expected revenue*. Auctioneer (Coordinated VO scheduler) collects bids from the bidding agents, and selects the winner(s) with the lowest bid. Afterwards, content is replicated there. The winning bidders are paid by the second-lowest bid and a VO is formed consisting of the buyer and seller CSDNs (winners) based on a common goal to replicate content, and to serve it to the end-users in an efficient manner.

The viability of a VO may change depending on the demand of content and the participant's economic gain. A VO participant (seller) should be able to adapt with the change in context and to vary their bidding valuation accordingly. Hence, re-negotiation should take place among the VO participants to either disband or rearrange the VO into a new organization that better fits the prevailing circumstances.

## 6.1. Payoff Function of the buyer

In this section we define the payoff function of the buyer. During auction, buyer uses the payoff function to internally determine the maximum payable amount that it can spend for replicating content $c$ at time $t_k$. It is denoted as $P_{\max}(c, t_k)$. Buyer takes replication decision by selecting the lowest bid(s) if and only if the bidding amount is less than or equal to $P_{\max}$. If the bid of a seller $i$ is greater than $P_{\max}$, it will not be accepted as it follows $B_i \leq P_{\max}$. In the worst case, auction terminates when the bid of all seller is greater than $P_{\max}$. In such case, buyer modifies its auction policy and the auctioneer reinitiates auction with changed auction policy.

When a buyer $i$ finds that it cannot handle the client request, i.e., $\partial(L_i, t_k) > 0$, it incurs cost (i.e. penalty) since it was incapable to provide service the users and to meet the service level objectives when required. We can quantify it as $\alpha \partial(L_i, t_k)$, where $\alpha$ is the penalty constant for the client requests that are not serviced by the buyer. Hence, maximum payable amount $P_{\max}$ for content $c$ at time $t_k$ is defined as:

$$P_{\max} = \beta \left[ \frac{\sum_{j=1}^{k-1} \{p_j + \gamma\} \delta(r_k^c, r_j^c) \phi(r_k^l, r_j^l) e^{-\lambda(t_k - t_j)}}{(k-1)} \right] - \alpha \partial(L_i, t_k)$$

The first term in the above equation is used to predict the amount to be paid for content $c$ at time $t_k$, taking past information as input. The second term in the equation deducts the cost incurred by the buyer (for taking replication decision) from the predicted amount to be paid. To make the prediction of amount to be paid, buyer exploits the similarity, distance and time of content requests. It considers the average of the weighted sum of the maximum amount paid previously for the content $c$ before time $t_k$. The values $\beta, \gamma, \lambda$ represent the constant coefficient values such that $\beta + \gamma + \lambda = 1$. The summation in the first term cycles over the content requests stored in the history log. As stated earlier, each content request $r$ is a tuple $\{r^c, r^l\}$, where $c$ denotes the content and $l$ denotes the location of the request. For each request the maximum payment in previous content request (plus a constant) is scaled according to a similarity function $\delta$, a distance function $\phi$, and an exponential time decay function.

The similarity function $\delta(c_k, c_j); c_k, c_j \in C$ is a measure of similarity between to content requests such that $0 \leq \delta(c_k, c_j) \leq 1$. For given two content requests $r_k$ and $r_j$, the similarity function is defined as $\delta(r_k^c, r_j^c)$. The distance function $\phi(l_k, l_j)$ takes into account the locality two content requests. It is constrained by $0 \leq \phi(l_k, l_j) \leq 1$. For given two content requests $r_k$ and $r_j$, the distance function is defined as $\phi(r_k^l, r_j^l)$. The exponential term in the equation causes older information to be weighted with lower importance than the more recent information.

## 6.2. Bidding and Utility Function of seller

The bidding agents of the Web servers bid with the amount as determined by the *Bidding Function* $B_i(c_k, t_k) = S_i(c_k) + UF_i(c_k, t_k) \pm \psi_i(B)$, where $S_i(c_k)$ is the storage cost incurred by seller $i$ to replicate $c_k$, $UF_i(c_k, t_k)$ is the *Utility Function*, and $\psi_i(B)$ is a function to reflect sellers interest in bidding. The storage cost $S_i(c_k)$ is defined as: $S_i(c_k) = \Re(c_k)\varphi(S_i)$, where $\Re(c_k)$ is the storage requirements of the buyer for content $c_k$ and $\varphi(S_i)$ is the unit cost of storage space on the seller $i$. Bidding agents uses a *Utility Function* $UF_i(c_k, t_k)$, which specifies the expected benefit by selling storage space to the buyer for replicating content $c_k$. Here, we assume that the agents are rational in the sense that they are trying to maximize their utility and they will not do an auction that yields them a negative utility. For replicating new content, seller may have to delete the old content stored in it. A seller does not bid for selling its storage space unless it is economically beneficial for it. Hence, the $UF$ for seller $i$ at time $t_k$ for content $c_k$ becomes: $UF_i(c_k, t_k) = ER_{new} - ER_{old}$. Where $ER_{new}$ and $ER_{old}$ respectively denote the expected revenue from current content and the existing content that might have to be deleted to make space for new content. In all the cases, successful replication yields $UF(c_k, t_k) > 0$. To predict the expected revenue for the content to be replicated, the seller depends on its history of requests for similar content, and user-request pattern of that content.

## 6.3. Prediction of revenue

The function for revenue calculation depends on the expected requests for particular content over a particular time period $T$ for which a request has been received at time $t_k$. To predicted the requests for content we first the model of content request pattern as symmetric random walk with variable step size. We then define the revenue prediction function according to Binomial distribution. Then, we also define the prediction function according to the popularity of content based on Zipf distribution.

### 6.3.1 Modeling content request based on random walk

Here we introduce the concept of content similarity as in [18]. We define *content space* $\{C\}$ as the set of all content requests from the end-users and *content-ID space* $\{c\}$ as a set of content-request identifiers. Content similarity is defined to be a mapping between $\{C\}$ and $\{c\}$. If two content requests are $c_1$ and $c_2$, the smaller the difference $|c_2 - c_1|$, the higher is the similarity between the corresponding contents. Based on these assumptions, the history of content requests can be represented as a *random walk* in the space $\{c\}$ of request identifiers. A random walk consists of a sequence $<c_i, i \geq 0>$ of identifiers, starting from $c_0$, adding a sequence $<s_i, i \geq 0>$ of random-walk steps, in which $s_i = c_i - c_{i-1}$ for $\forall i > 0$. Each generic step $s$ is an independent random variable. Hence, the content request pattern can be modeled as a random walk consisting of a sequence of variable step size. Figure 3 shows an example random walk. Here, content request starts from $c_0$ and covers $n$ requests over time $t_0 \sim t_n$.

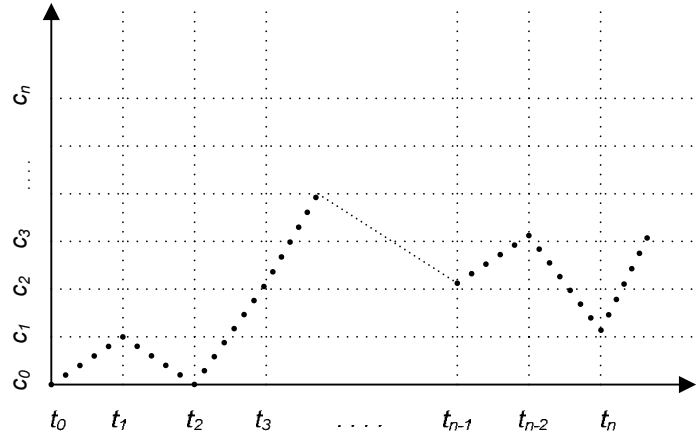

**Figure 4: Content request pattern as a random walk in request identifier space** $\{r\}$

**Prediction of revenue from content:** Now we define a function that calculates the expected revenue based on predicted requests for a particular content over a particular time period $T$ for which a request has been received at time $t_k$. This function depends on the request pattern of next $n$ content requests, and on the history of request for similar content. The buyer provides the history of content requests during auction. Hence, the function for determining expected revenue is defined as:

$$ER(c_k, t_k, n) = \rho \sum_{j=k+1}^{k+n} \delta(r_k^c, r_j^c) \phi(r_k^l, r_j^l) + (1-\rho) \sum_{i=1}^{k-1} \delta(r_k^c, r_i^c) \phi(r_k^l, r_i^l)$$

The function $ER(c_k, t_k, n)$ calculates the expected revenue for content $c_k$ at time $t_k$, by taking into account the request pattern of next $n$ content requests, and the history for similar content request as provided by the buyer. $\delta(c_i, c_j)$ is the similarity function that exploits the closeness of two content requests and is constrained by $0 \leq \delta(c_i, c_j) \leq 1$. $\phi(r_i, r_j)$ is the distance function which is constrained by $0 \leq \phi(r_i, r_j) \leq 1$. It determines the distance between the present and previous content requests and thus exploits the locality of content requests. The second term in the above equation manipulates the history of similar content requests in the past. The constant $\rho, 0 \leq \rho \leq 1$ works as an 'impact factor' on the predicted content requests and on the history of content request.

### 6.3.2   Defining prediction function based on Binomial distribution

Our goal is now to obtain a simplified and explicit form of the function $ER(C_k, t_k, n)$. In order to perform calculations and to get an explicit form of $ER(C_k, t_k, n)$ we choose a particular probability distribution of generic step size $s$, where $s$ being the step between two successive content request identifiers in the random walk. We consider that the generic step size $s$ in the random walk is a discrete random variable with binomial distribution. We also assume that such distribution for integer random step size is symmetric and centered to zero. Each content request $c_k$ represents the content requested at $k$-th step of the random walk. It can be represented as: $c_k = c_0 + \sum_{i=1}^{k} s_i$, since the content request starts from $c_0$ and contains first $k$ steps of $<s_i, i \geq 0>$. Since each $s_i$ is an independent random variable with symmetric binomial distribution, each generic $c$ is also a random variable with the same distribution.

Hence, the probability of receiving a content request $c_k$ at step $k$ is given by,

$$P_k(c_k) = \frac{1}{2^{2kS}} \binom{2kS}{c_k - \bar{s} + kS}, \quad |c_k - \bar{s}| \leq kS;$$

where $\bar{s}$ is the mean value of the binomial distribution, $S$ is the maximum value for an integer random step in the range $[-S, S]$, $c_k$ is the identifier for content requested at $k$-th step of the random walk. Summing up all the terms $P_k(c_k)$ for k going from $1$ to $n$, we obtain the totkal number of times a content $c_k$ is requested during the next $n$ requests. So, the expected revenue $ER(c_k, t_k, n)$ for content $c_k$ during the next $n$ requests according to a binomial distribution can be defined as: $ER(c_k, t_k, n) = \sum_{i=1}^{n} P_i(c_k)$.

If $(k-1)$ is the number of previous requests during time interval $T'$ in the past which serves as the basis for our estimation, the number $n$ of next content requests during future time interval $T$ is given by, $n = (k-1)\frac{T}{T'}$. Here, we assume that the mean arrival rate of the content requests is constant. The value of $S$ depends on the number of previous requests $r$ and the mean value $\bar{s}$ is calculated as a weighted average of the last $(k-1)$ steps, where weights decrease as going towards past time.

### 6.3.3 Defining prediction function based on Zipf distribution

Requests for Web content in the Internet follow Zipf's law [19]. Zipf's law state that the relative probability of a request for the $k$-th most popular content is proportional to $1/k$. It has been observed that the requests for Web content from fixed group of users follow a Zipf-like distribution, $\Omega/k^\mu$ with $0 < \mu \leq 1$ [20]. The value of $\mu$ depends on the concentration of requests for hot contents. Now, the cumulative probability of receiving a request for content $c_k$ at time $t_k$ is given by, $\widetilde{P}_k(c_k) = \sum_k \Omega/c_k^\mu \approx \Omega c_k^{1-\mu}/(1-\mu)$. Here, $\Omega$ is a normalizing constant defined as, $\Omega = (1-\mu)/C^{1-\mu}$, where total number of content is denoted as $C$. Hence, we get $\widetilde{P}_k(c_k) \approx (c_k/C)^{1-\mu}$. Since probability increases with larger values of $\mu$, it indicates that requests are concentrated on hot contents. Now the expected revenue for content $c_k$ is for $n$ requests is given by, $ER(c_k, t_k, n) = \sum_{i=1}^{n} \widetilde{P}_i(c_k)$.

## 7. Summary and Future Works

In this paper, we have presented an open, scalable and Service-Oriented Architecture (SOA)-based system to assist the creation of open Content and Service Delivery Networks according to a VO model. To encourage resource sharing and peering arrangements among different CDN providers at global level, we propose the use of market models. Hence, we introduce an economy-based content replication strategy based on auction protocol for replicating content in surrogates of peering CDNs. The use of economic concepts in this context provides a solid basis for rational agents in peering CSDN environment to decide whether to attend in peering constellation. W

The proposed content replication mechanism in peering CDN environment should be effective as it adapts with the dynamism of the system. It is also be able to take replication decision on-demand in a distributed manner with the change in content behavior, and thus reduce user-perceived latency. To the best of our knowledge, none of the works performed in the content internetworking domain have explored the emergent marketplace behavior of such systems using economic concepts. Use of the economic model may be the basis of a replication mechanism that dynamically adapts to the changes in content request pattern, and make decision to replicate content to the surrogates of peering CSDNs in areas which exhibits the potential to generate Web hotspots. We expect that, realizing our VO model for forming CSDNs and the policy framework within the VO should be a timely contribution to the ongoing content-networking trend.

We are currently refining the design of the described architecture and also building a prototype system based on the models presented in the paper. In addition to this work we plan to address other related issues for content replication and load distribution among peering CDNs. This includes predicting revenue from replication considering the content request patterns according to popularity distribution. For more information, please visit the project Web site at www.gridbus.org/cdn.

## Acknowledgements

We are thankful to Kyong Hoon Kim of the University of Melbourne for sharing thoughts on this topic, and for discussion regarding the analytical formulation of system model.